\documentclass{PoS}

\newif\ifpdf\ifx\pdfoutput\undefined\pdffalse\else\pdfoutput=1\pdftrue\fi

\def\be{\begin{equation}}
\def\ee{\end{equation}}

\def\bea {\begin{eqnarray}}
\def\eea {\end{eqnarray}}

\title{Theory update on the inlcusive determination of $|V_{cb}|$}

\ShortTitle{Inclusive determination of $|V_{cb}|$}

\author{\speaker{Soumitra Nandi}\thanks{I would like to thank Andrea Alberti, Thorsten Ewerth, Paolo Gambino 
          and Kristopher Healey for fruitful collaboration}\\
        \sl Department of Physics, Indian Institute of Technology, Guwahati 781039, India \\
        E-mail: \email{soumitra.nandi@iitg.ernet.in}}

\abstract{ In this talk I update the precision determination of the CKM element $|V_{cb}|$ from the inclusive decay 
$B\to X_c\ell\nu_{\ell}$. }

\FullConference{CKM, 2016}

\begin{document}

\section{Motivation}
One of the primary goals of the study of $B$  meson decays and mixing is to construct the unitarity triangle (UT), 
which is defined by the relation $V_{cd}V^{\ast}_{cb} + V_{ud}V^{\ast}_{ub} + V_{td}V^{\ast}_{tb}=0$. 
The Cabbibo-Kobayashi-Masakawa (CKM) elements $V_{ij}$ ($i= u,c,t$ and $j = d,s,b$) are parametrized in terms of four 
independent parameters: $\eta$, $\rho$, $\lambda$ and A. In order to construct the UT, various measurements which are 
sensitive to the these CKM elements are projected into the ($\rho$, $\eta$) plane. The precise measurements of 
$V_{ub}$ and $V_{cb}$ are important since they play an important role in finding out the regions in the 
$(\rho, \eta)$ plane where the apex of the triangle should lie. In this regard, the loop induced and CP violating $B$ 
decays are also important. For details, see \cite{Battaglia:2003in}. 

The tree level semileptonic decays $b\to c\ell\nu_{\ell}$ ($\ell = e, \mu$) are crucial for the determination of $V_{cb}$. 
It can be extracted from both exclusive decays, like $B\to D^{(\ast)}\ell\nu$, and inclusive decays, like 
$B \to X_c \ell\nu_{\ell}$. These decays are expected to be free from any new physics (NP)  effects, hence 
provide a clean environment for the measurement of the $V_{cb}$. The inclusive channels are relatively clean, and the 
decay rates have a solid description via operator product expansion (OPE) or heavy quark expansion (HQE) \cite{OPE}. 
In these decays, the non perturbative unknowns can be extracted using the final state lepton and hadron energy distribution 
\cite{Gambino:2004,Gambino:2013}. These are also useful to extract the $b$-quark masses, and for a consistency check of the OPE/HQE and other effective theory
approaches. As per the measurement is concerned, it has small statistical and systematic errors, and highly 
sensitive to the theoretical uncertainties, for details, see \cite{hfag}. Therefore, precise predictions in the SM 
including reliable uncertainties are possible. 

The exclusive semileptonic decays have similar solid 
descriptions in terms of heavy quark effective theory (HQET) \cite{exclusive}.  On contrary to the inclusive decays, 
the non-perturbative unknowns in the exclusive decays can not be extracted experimentally. One needs to calculate 
them and that is where the major challenges lie. 

\section{Framework}

The decay rate distribution for the decay $B\to X_c\ell\nu_{\ell}$ is given by  
\begin{equation}
\frac{d\Gamma}{dq^2 dE_e dE_\nu} = 2 G_F^2 V_{cb}^2 W_{\mu\nu} L^{\mu\nu}, 
\end{equation}
where $L^{\mu\nu}$ and $W_{\mu\nu}$ are the leptonic and hadronic tensors respectively, and they are defined as
\begin{equation}
 L^{\mu\nu} = 2 \big(p_e^\mu p_{\bar\nu}^\nu + p_e^\nu p_{\bar\nu}^\mu - g^{\mu\nu} p_e. p_{\bar\nu} - i 
 \epsilon^{\eta\nu\lambda\mu} (p_e)_\eta (p_{\bar\nu})_\lambda \big),
\end{equation}
\begin{equation}
 W^{\mu\nu} = \frac{1}{2 m_B} \sum_X (2\pi)^3\delta^4(p_B - q - p_X)\times \langle  B (p_B) | J^{\dag\mu}_L| 
 X_c(p_X)\rangle \langle X_c(p_X) | J^{\nu}_L|  B (p_B)\rangle,
\end{equation}
with $J_L^{\mu} = {\bar c}\gamma^{\mu} P_L b $, for details see the reviews \cite{hfag,pdg,Bevan:2014} and the references 
therein.

Here, the major theoretical challenges are the calculation of the tensor $W_{\mu\nu}$ which represents 
the hadronic contribution to the decay width. Using the optical theorem,
the hadronic tensor can be calculated from the imaginary part of the forward scattering amplitude, 
\begin{equation}
W_{\mu\nu} \propto Im(T_{\mu\nu}),
\end{equation}    
where $T_{\mu\nu}$ is defined as the forward matrix element of the time ordered product of the two currents,
\begin{equation}
 T^{\mu\nu} = - i \int d^4x e^{-i q.x} \frac{\langle  B| T[ J_L^{\mu\dag}(x) J_L^{\nu}(0)]| B \rangle}{2 m_B}.
\label{tmunu}
 \end{equation} 
If the energy released in the decays of $b$-quark mediated by weak interactions is large, then such decays will take place on 
a time scale which is much shorter than the time it takes the quarks in the final state to form physical hadronic states. 
Hence, the inclusive decay rates may be modeled simply by the decay of free $b$ quark. 
Also, since the energy released in such decays are much larger than the hadronic scale,
they are largely insensitive to the details of the initial hadronic structure. 
This intuitive picture is formalized by OPE. In the limit $M_W >> m_b >> \Lambda_{QCD}$, we can organize an expansion 
in $\Lambda_{QCD}/m_b$, with the leading term corresponding to the free quark decay. 
Therefore, the right hand side of eq. \ref{tmunu} can be written as an infinite sum of local operators ($O_i$)
of increasing dimension

\begin{figure}[htbp]
\includegraphics[width=1.0\textwidth]{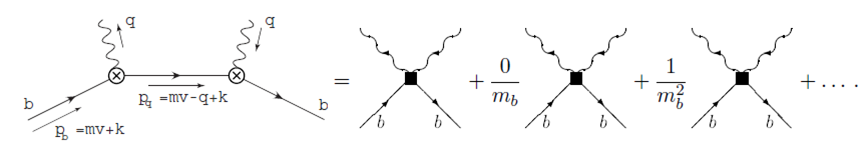}
\caption{Tree level matching conditions}
\label{matching}
\end{figure}
\begin{equation}
 - i \int{d^4x e^{-iq.x} T[J^{\mu\dag}_L(x) J^\nu_L(0)]} = \sum_i C_i O_i , 
\end{equation}
where the Wilson coefficients $C_i$ can be expressed as perturbative series in $\alpha_s$.   
The lowest dimensional term will dominate in the limit $m_b \to \infty$. These Wilson coefficients at tree level and at 
the loop level can be obtained from the matching conditions. As an example, the tree level matching is diagrammatically
shown in Fig. \ref{matching} with $q = c$. In order to get matching conditions at ${\cal O}(\alpha_s)$ diagrams with 
one loop need to be considered. The two-index amplitude $T^{\mu\nu}$ can be decomposed into five tensor structures  
\begin{equation}
 T^{\mu\nu} = - g^{\mu\nu} T_1 + v^{\mu}v^{\nu} T_2 - i \epsilon^{\mu\nu\alpha\beta} v_{\alpha} q_{\beta} T_3 
 + q^{\mu}q^{\nu} T_4 + (v^{\mu}v^{\nu} + v^{\nu}v^{\mu}) T_5 ,
\end{equation}
where the $T_i$s are known as the projectors. They are the functions of Lorentz invariant quantities $q^2$ and $q.v$, where 
$v = P_B/M_B$ is the four velocity of the decaying $B$ meson.  
These projectors can be expressed as a double series expansion: perturbative in $\alpha_s$ and 
non-perturbative in $\Lambda_{QCD}/m_b$, such as 
\begin{equation}
 T_i = \sum_{n \ge 3} \sum_{j \ge 0} \left(\frac{\Lambda_{QCD}}{m_b} \right)^{n-3} \left(\frac{\alpha_s}{\pi} \right)^j
 c_{ij}^{(n)} \langle { B}| O^{n} | {B} \rangle  .  
 \end{equation}

Therefore, the decay width for $B \to X_c \ell \nu_{\ell}$ can be written as \cite{Benson:2003}
\begin{eqnarray}
 \Gamma_{SL} &=& |V_{cb}|^2 \frac{G_F^2 m_b^5}{192 \pi^3} ( 1 + A_{EW}) \times \left[C_0^0 + \frac{0}{m_b} 
 + C_2(r, \frac{\mu_{\pi}^2}{m_b^2}, \frac{\mu_{G}^2}{m_b^2}) + C_3(r, \frac{\rho_{D}^3}{m_b^3}, \frac{\rho_{LS}^3}{m_b^3})
 + .... \right. \nonumber \\
&{}&\left. + .....+ \alpha_s\left( C_0^1 + C_2^1(r, \frac{\mu_{\pi}^2}{m_b^2}, \frac{\mu_{G}^2}{m_b^2}) +  
 C_3^1(r, \frac{\rho_{D}^3}{m_b^3}, \frac{\rho_{LS}^3}{m_b^3}) + ... \right) + ... \right],
\end{eqnarray}
where $r = m_c/m_b$. Here, $A_{EW}$ represents the electroweak corrections the ultraviolet renormalization of the
Fermi interaction, and $1+ A_{EW} \approx 1.014 $ \cite{sirlin}. The $C_i^{j}$s are the perturbatively 
calculable functions. The parameters like $\mu_\pi^2$, $\mu_G^2$, and  $\rho_D^3$, $\rho_{LS}^3$ are 
the matrix elements of the non-perturbative operators at order $(\Lambda/m_b)^2$ (kinetic and chromomagnetic ) and 
$(\Lambda/m_b)^3$ (Darwin and spin-orbit) respectively. They are defined as given below
\begin{equation}
m_B \mu_\pi^2 \sim \langle  B | {\bar b}_v D^{\mu} D_{\mu} b_v | B \rangle , ~~~~~~~~~~~~~~ 
 m_B \mu_G^2 \sim \langle B | {\bar b}_v g_s G_{\mu\nu} \sigma^{\mu\nu} b_v | B \rangle  , 
 \label{dim5opt}
\end{equation}
and 
\begin{eqnarray}
 m_B \rho_D^3 &\sim  & \langle  B | {\bar b}_v \left[ iD_{\mu}, [i D_{\sigma}, i D_{\nu}]\right] b_v | B \rangle ~\Pi^{\mu\nu}
 v^\sigma , \nonumber \\ 
 m_B \rho_{LS}^3 &\sim & \langle  B | {\bar b}_v \left\{ iD_{\mu}, [i D_{\sigma}, i D_{\nu}]\right\} (-i \sigma_{\alpha\beta}) 
 b_v | B \rangle~ \Pi^{\alpha\mu} \Pi^{\beta\nu} v^\sigma,  
 \label{dim6opt}
\end{eqnarray}
where $\Pi_{\mu\nu} = (g_{\mu\nu} - v_\mu v_\nu)$. We note that the decay rate is sensitive to the CKM element $|V_{cb}|$. 
Also, the main sources of uncertainties are: (i) mass of the $b$ quark and the ratio $r$, (ii) higher 
order QED and QCD radiative corrections, (iii) higher order of the $1/m_b$ corrections, (iv) extractions 
of HQE parameters, and (v) parton hadron duality \cite{duality}. 

The OPE/HQE parameters can be extracted from the moments of the differential distributions, like the leptonic 
energy moments are defined as \cite{Gambino:2013}

\begin{equation}
 M_1^{\ell} = \frac{1}{\Gamma}\int dE_{\ell} E_{\ell} \frac{d\Gamma}{dE_{\ell}}; ~~~~~~~
 M_n^{\ell} = \frac{1}{\Gamma}\int dE_{\ell} ( E_{\ell} - M_1^{\ell})^n \frac{d\Gamma}{dE_{\ell}}~~(n >1).
\end{equation}
with 
\begin{equation}
 M_n^{\ell} = (\frac{m_b}{2})^n \left[\phi_n(r) + {\bar a}_n(r) \frac{\alpha_s}{\pi} + {\bar b}_n(r) \frac{\mu_\pi^2}{m_b^2}
 + {\bar c}_n(r) \frac{\mu_G^2}{m_b^2} + {\bar d}_n(r) \frac{\rho_D^3}{m_b^3} + {\bar s}_n(r) \frac{\rho_{LS}^3}{m_b^3}+ ...
 \right].
\end{equation}
Similarly, the moments of the hadronic mass is given by
\begin{equation}
 M_1^{X} = \frac{1}{\Gamma}\int dM_X^2 (M_X^2 - {\bar M}_D^2 ) \frac{d\Gamma}{dM_X^2}; ~~~~~~~
 M_n^{X} = \frac{1}{\Gamma}\int dM_X^2 (M_X^2 - \langle M_X^2 \rangle )^n \frac{d\Gamma}{dM_X^2}~~(n >1).
\end{equation}
with 
\begin{equation}
 M_n^{X} = m_b^{2n}\sum_{l=0} \left[\frac{M_B -m_b}{m_b}\right]^l \left[ E_{nl}(r) + 
 {\bar a}_{nl}(r) \frac{\alpha_s}{\pi} + {\bar b}_{nl}(r) \frac{\mu_\pi^2}{m_b^2}
 + {\bar c}_{nl}(r) \frac{\mu_G^2}{m_b^2} + {\bar d}_{nl}(r) \frac{\rho_D^3}{m_b^3} + {\bar s}_{nl}(r) 
 \frac{\rho_{LS}^3}{m_b^3}+ ... \right].
\end{equation}
Instead of the linear moments, which are highly correlated, the central moments are more useful in the extraction of  
physical information. Here, the functions $\phi_n(r)$, ${\bar a}_{n(l)}(r)$, ${\bar b}_{n(l)}(r)$, ${\bar c}_{n(l)}(r)$, 
 ${\bar d}_{n(l)}(r)$, ${\bar s}_{n(l)}(r)$ are calculable perturbatively. We note that these moments are highly sensitive to the quark masses and 
 the OPE parameters. Therefore, a global fit to decay rate and moments allows us to extract $|V_{cb}|$, $m_b$, $m_c$, $\mu_\pi^2$, 
$\mu_G^2$, $\rho_D^3$ and $\rho_{LS}^3$. 

\section{Measurements of $|V_{cb}|$: State of the art}
So far a lot of progress has been made in improving the precision in the extractions of $V_{cb}$ and the OPE parameters. 
At tree level, i.e at leading order in $\alpha_s$, all the corrections up to order $1/m_b^5$ have been estimated 
\cite{Gremm:1996,Mannel:2010}. A  large number of parameters are associated with the ${\cal O}(1/m_b^{4,5})$ corrections, and 
hence, they can not be fitted directly from the experimental data. Therefore, these parameters are estimated using ground 
state saturation approximation. Only the parameters relevant up to the ${\cal O}(1/m_b^{2,3})$ are fitted directly from 
the experimental data. The corrections to the partonic rate and leptonic and hadronic mass distributions are fully known 
at order $\alpha_s$ \cite{ Jezabek} and ${\cal O}( \alpha_s^2 \beta_0)$ \cite{Gremm}. The two loop corrections of order 
${\cal O}(\alpha_s^2)$ to the width and first few moments are calculated \cite{Pak:2008}. The ${\cal O}(\alpha_s)$ 
corrections to the kinetic operator have been estimated only numerically in \cite{Becher:2007}. In the last few years, 
a complete analytical calculation of the ${\cal O}(\alpha_s \Lambda_{QCD}^2/m_b^2)$ corrections have been performed. 
For details, see the references \cite{Alberti:2012}. Similar calculations to the Wilson coefficients of the 
dimension-6 operators, defined in \ref{dim6opt}, are ongoing \cite{Alberti:2017}. 

After incorporating all the known corrections to the decay width and moments, and fitting all the relevant parameters with 
the available data \cite{experiments} on width and moments, one obtains \cite{Gambino:2016}, 
\begin{equation}
 \frac{\Gamma}{z(r) \Gamma_0} = 1 - 0.116_{\alpha_s} - 0.030_{{\alpha_s}^2} - 0.042_{1/m_b^2} - 0.002_{\alpha_s/m_b^2} 
 - 0.030_{1/m_b^3} + 0.005_{1/m_b^4} + 0.005_{1/m_b^5} .
\label{allcorr}
 \end{equation}
In this expression $z(r) = 1 - 8 r + 8 r^3 - r^4 - 12 r^2 ln(r)$ and $\Gamma_0 =(1+ A_{EW}) |V_{cb}|^2 G_F^2 m_b^5/{192\pi^3}$.
We note that the width depends on the fifth power of the mass of $b$ quark. Hence, the uncertainties associated with this 
mass has great impact in the precision extractions of $|V_{cb}|$. In the fitting, the $b$ quark mass and the 
non-perturbative matrix elements are expressed in the kinetic scheme \cite{Gambino:2004}, setting the cutoff $\mu^{kin}$ 
at 1 {\it GeV}. Details of the fitting procedure and various inputs can be seen from ref. \cite{Gambino:2013}. 
The extracted value of $V_{cb}$ without incorporating the recently calculated $\alpha_s/m_b^2$ and $1/m_b^{4,5}$ 
corrections is given by \cite{Gambino:2013}
\begin{equation}
 |V_{cb}| = (42.42 \pm 0.86)\times 10^{-3},
\end{equation}
the estimated error is $\approx$ 2\%. After the inclusion of $\alpha_s/m_b^2$ corrections the value is \cite{Alberti:2014} 
\begin{equation}
 |V_{cb}| = (42.21 \pm 0.78)\times 10^{-3}. 
\end{equation}
The error is reduced to 1.8\% and the central value is reduced by 5\% too.  Including all the known corrections given in 
eq. \ref{allcorr}, one obtains \cite{Gambino:2016}
\begin{equation}
 |V_{cb}| = (42.11 \pm 0.74)\times 10^{-3}.
\end{equation}
Here we also note that the central value has reduced by only 0.25\% after the inclusion of ${\cal O}(1/m_b^{4,5})$ effects, and the 
estimated error is 1.7\%.

\section{Conclusions}

The onset of Belle-II experiment will bring us to a high precision era. Considerable progress has been made towards improving
the precision $|V_{cb}|$. A more precise extraction of $|V_{cb}|$ is necessary in order to understand the SM, QCD 
approaches, and for an implicit search of NP. There is much more to do in order to improve the precision. 

\section{Acknowledgement}

I would like to thank the organizers of CKM 2017 for inviting me to present the updates on inclusive determination of
$|V_{cb}$|.


\begin{thebibliography}{99}
\bibitem{Battaglia:2003in} 
  M.~Battaglia {\it et al.},
  hep-ph/0304132.

\bibitem{OPE}
  I.~I.~Y.~Bigi, M.~A.~Shifman, N.~Uraltsev and A.~I.~Vainshtein,
  Phys.\ Rev.\ D {\bf 59}, 054011 (1999)
  [hep-ph/9805241];
  I.~I.~Y.~Bigi, M.~A.~Shifman and N.~Uraltsev,
  Ann.\ Rev.\ Nucl.\ Part.\ Sci.\  {\bf 47}, 591 (1997)
  [hep-ph/9703290];
  I.~I.~Y.~Bigi, B.~Blok, M.~A.~Shifman, N.~Uraltsev and A.~I.~Vainshtein,
  [hep-ph/9401298];
  I.~I.~Y.~Bigi, M.~A.~Shifman, N.~G.~Uraltsev and A.~I.~Vainshtein,
  Int.\ J.\ Mod.\ Phys.\ A {\bf 9}, 2467 (1994)
  [hep-ph/9312359];
  N.~Uraltsev,
  [hep-ph/0010328].
  \bibitem{Gambino:2004} 
  P.~Gambino and N.~Uraltsev,
  Eur.\ Phys.\ J.\ C {\bf 34}, 181 (2004)
  [hep-ph/0401063].
  
 \bibitem{Gambino:2013} 
  P.~Gambino and C.~Schwanda,
  Phys.\ Rev.\ D {\bf 89}, no. 1, 014022 (2014)
  [arXiv:1307.4551 [hep-ph]].
 
\bibitem{hfag}
  Y.~Amhis {\it et al.},
  arXiv:1612.07233 [hep-ex].
 
\bibitem{exclusive} 
  N.~Isgur and M.~B.~Wise,
  Phys.\ Lett.\ B {\bf 232}, 113 (1989).
  M.~A.~Shifman and M.~B.~Voloshin,
  Sov.\ J.\ Nucl.\ Phys.\  {\bf 47}, 511 (1988)
  [Yad.\ Fiz.\  {\bf 47}, 801 (1988)].
  J.~A.~Bailey {\it et al.} [MILC Collaboration],
  Phys.\ Rev.\ D {\bf 92}, no. 3, 034506 (2015)
  [arXiv:1503.07237 [hep-lat]].
  
 \bibitem{pdg} 
  C.~Patrignani {\it et al.} [Particle Data Group],
  Chin.\ Phys.\ C {\bf 40}, no. 10, 100001 (2016).
 

 
\bibitem{Bevan:2014} 
  A.~J.~Bevan {\it et al.} [BaBar and Belle Collaborations],
  Eur.\ Phys.\ J.\ C {\bf 74}, 3026 (2014)
  doi:10.1140/epjc/s10052-014-3026-9
  [arXiv:1406.6311 [hep-ex]].
 
\bibitem{Benson:2003} 
  D.~Benson, I.~I.~Bigi, T.~Mannel and N.~Uraltsev,
  Nucl.\ Phys.\ B {\bf 665}, 367 (2003)
  [hep-ph/0302262].
  
\bibitem{sirlin}
A.~Sirlin, 
Nucl.\ Phys.\ B {\bf 71} (1974) 29-51

\bibitem{duality} 
  I.~I.~Y.~Bigi and N.~Uraltsev,
  Int.\ J.\ Mod.\ Phys.\ A {\bf 16}, 5201 (2001)
  doi:10.1142/S0217751X01005535
  [hep-ph/0106346].

\bibitem{Gremm:1996} 
  M.~Gremm and A.~Kapustin,
  Phys.\ Rev.\ D {\bf 55}, 6924 (1997)
  [hep-ph/9603448].
\bibitem{Mannel:2010} 
  T.~Mannel, S.~Turczyk and N.~Uraltsev,
  JHEP {\bf 1011}, 109 (2010)
  [arXiv:1009.4622 [hep-ph]].
\bibitem{Jezabek} 
  M.~Jezabek and J.~H.~Kuhn,
  Nucl.\ Phys.\ B {\bf 320}, 20 (1989);
  M.~Jezabek and J.~H.~Kuhn,
  Nucl.\ Phys.\ B {\bf 314}, 1 (1989);
  A.~Czarnecki and M.~Jezabek,
  Nucl.\ Phys.\ B {\bf 427}, 3 (1994)
  [hep-ph/9402326];
  A.~F.~Falk, M.~E.~Luke and M.~J.~Savage,
  Phys.\ Rev.\ D {\bf 53}, 2491 (1996)
  [hep-ph/9507284];
  A.~F.~Falk and M.~E.~Luke,
  Phys.\ Rev.\ D {\bf 57}, 424 (1998)
  [hep-ph/9708327];
  M.~Trott,
  Phys.\ Rev.\ D {\bf 70}, 073003 (2004)
  [hep-ph/0402120].
  
  
\bibitem{Gremm} 
  M.~Gremm and I.~W.~Stewart,
  Phys.\ Rev.\ D {\bf 55}, 1226 (1997)
  [hep-ph/9609341];
  V.~Aquila, P.~Gambino, G.~Ridolfi and N.~Uraltsev,
  Nucl.\ Phys.\ B {\bf 719}, 77 (2005)
  [hep-ph/0503083].

\bibitem{Pak:2008} 
A.~Pak and A.~Czarnecki,
  Phys.\ Rev.\ Lett.\  {\bf 100}, 241807 (2008)
  [arXiv:0803.0960 [hep-ph]];
K.~Melnikov,
  Phys.\ Lett.\ B {\bf 666}, 336 (2008)
  [arXiv:0803.0951 [hep-ph]];
  S.~Biswas and K.~Melnikov,
  JHEP {\bf 1002}, 089 (2010)
  [arXiv:0911.4142 [hep-ph]];
  P.~Gambino,
  JHEP {\bf 1109}, 055 (2011)
  [arXiv:1107.3100 [hep-ph]].

  \bibitem{Becher:2007} 
  T.~Becher, H.~Boos and E.~Lunghi,
  JHEP {\bf 0712}, 062 (2007)
  [arXiv:0708.0855 [hep-ph]].
  
 \bibitem{Alberti:2012} 
  A.~Alberti, T.~Ewerth, P.~Gambino and S.~Nandi,
  Nucl.\ Phys.\ B {\bf 870}, 16 (2013)
  [arXiv:1212.5082 [hep-ph]];
  A.~Alberti, P.~Gambino and S.~Nandi,
  JHEP {\bf 1401}, 147 (2014)
  [arXiv:1311.7381 [hep-ph]];
  T.~Mannel, A.~A.~Pivovarov and D.~Rosenthal,
  Phys.\ Rev.\ D {\bf 92}, no. 5, 054025 (2015)
  [arXiv:1506.08167 [hep-ph]].


\bibitem{Alberti:2017}
A.~Alberti, T.~Ewerth, P.~Gambino and S.~Nandi,
in preparation.

\bibitem{experiments} 
  B.~Aubert {\it et al.} [BaBar Collaboration],
  Phys.\ Rev.\ D {\bf 81}, 032003 (2010)
  [arXiv:0908.0415 [hep-ex]];
  B.~Aubert {\it et al.} [BaBar Collaboration],
  Phys.\ Rev.\ D {\bf 81}, 032003 (2010)
  [arXiv:0908.0415 [hep-ex]];
  B.~Aubert {\it et al.} [BaBar Collaboration],
  Phys.\ Rev.\ D {\bf 81}, 032003 (2010)
  [arXiv:0908.0415 [hep-ex]];
  B.~Aubert {\it et al.} [BaBar Collaboration],
  Phys.\ Rev.\ D {\bf 81}, 032003 (2010)
  [arXiv:0908.0415 [hep-ex]];
  B.~Aubert {\it et al.} [BaBar Collaboration],
  Phys.\ Rev.\ D {\bf 69}, 111104 (2004)
  [hep-ex/0403030].
  S.~E.~Csorna {\it et al.} [CLEO Collaboration],
  Phys.\ Rev.\ D {\bf 70}, 032002 (2004)
  [hep-ex/0403052].
  A.~H.~Mahmood {\it et al.} [CLEO Collaboration],
  Phys.\ Rev.\ D {\bf 70}, 032003 (2004)
  [hep-ex/0403053].

\bibitem{Gambino:2016} 
  P.~Gambino, K.~J.~Healey and S.~Turczyk,
  Phys.\ Lett.\ B {\bf 763}, 60 (2016)
  [arXiv:1606.06174 [hep-ph]].
 
 
\bibitem{Alberti:2014} 
  A.~Alberti, P.~Gambino, K.~J.~Healey and S.~Nandi,
  Phys.\ Rev.\ Lett.\  {\bf 114}, no. 6, 061802 (2015)
  [arXiv:1411.6560 [hep-ph]].

  

 
 \end{thebibliography}
\end{document}